\documentclass[lettersize,journal]{IEEEtran}
\usepackage{amsmath,amsfonts}
\usepackage{algorithm}
\usepackage{array}
\usepackage[caption=false,font=normalsize,labelfont=sf,textfont=sf]{subfig}
\usepackage{textcomp}
\usepackage{stfloats}
\usepackage{url}
\usepackage{verbatim}
\usepackage{graphicx}

\usepackage{svg}
\usepackage{enumitem}
\usepackage{soul}

\usepackage{algpseudocode}
\usepackage[style=ieee,backend=biber]{biblatex}

\usepackage{eurosym}
\usepackage{amssymb} 
\usepackage[normalem]{ulem}
\usepackage{booktabs}

\usepackage{multirow}
\usepackage{makecell}
\usepackage{bm}
\usepackage{xcolor}
\usepackage[table]{xcolor}

\usepackage[colorlinks=true,
            linkcolor=black,
            urlcolor=customorange,
            citecolor=customorange]{hyperref}
\usepackage{cleveref}  
\definecolor{orange1}{RGB}{255,240,210} 
\definecolor{DarkOrange}{HTML}{778A4F}
\definecolor{orange2}{RGB}{255,150,70} 
\definecolor{customorange}{HTML}{FF6F00}
\definecolor{customlight}{HTML}{D0CCBE}
\definecolor{customgray}{HTML}{D9D9C1}
\definecolor{customwhite}{HTML}{FFFFFF}

\begin{document}

\title{OrderFusion: Encoding Orderbook for End-to-End Probabilistic \\ Intraday Electricity Price Forecasting}

\author{Runyao Yu$^{1,2}$, 
Yuchen Tao$^3$, 
Fabian Leimgruber$^2$, 
Tara Esterl$^2$, 
Jochen Stiasny$^1$,  \\
Derek W. Bunn$^4$, 
Qingsong Wen$^{5,6}$,  
Hongye Guo$^7$, 
Jochen L. Cremer$^{1,2}$
\\ $^1$Delft University of Technology, $^2$Austrian Institute of Technology, $^3$RWTH Aachen, \\ $^4$London Business School, $^5$University of Oxford, $^6$Squirrel AI,  $^7$Tsinghua University}

\maketitle
\begin{abstract}
Probabilistic intraday electricity price forecasting is becoming increasingly important for short-term power-system operation. With increasing renewable generation, demand-side flexibility, and storage assets, market participants need to adjust their positions under uncertainty closer to delivery.
Continuous intraday (CID) markets support this process by providing updated price signals, helping participants manage imbalance exposure and operational risk.
Unlike auction markets, CID trading in many jurisdictions is characterized by the continuous posting of buy and sell orders. This dynamic orderbook microstructure of price formation presents special challenges for price forecasting. Conventional methods represent the orderbook via domain features aggregated from buy and sell trades, or by treating it as a multivariate time series, but such representations neglect the full buy-sell interaction structure of the orderbook. This research therefore develops a new order fusion methodology, which is an end-to-end and parameter-efficient probabilistic forecasting model that learns a interaction-aware representation of the buy-sell dynamics. Furthermore, as quantile crossing is often a problem in probabilistic forecasting, this approach hierarchically estimates the quantiles with non-crossing constraints. Extensive experiments on CID price indices across high- and low-liquidity European markets demonstrate consistent improvements over conventional baselines, and ablation studies highlight the contributions of the main components.
The methodology is available at: \url{https://runyao-yu.github.io/OrderFusion/}.
\end{abstract}


\section{Introduction}
\label{introduction}
Whilst the wholesale market arrangements for electricity vary across many jurisdictions, a common approach follows a voluntary, self-dispatch competitive design based upon forward trading between market participants and subsequent real-time balancing for each delivery period, as administered by the system operator. 
The day-ahead auction tends to be the most liquid and widely-referenced forward product~\cite{nickelsen2025bayesian}. 
Ex post, a settlement agent then typically clears the 
imbalances between forward commitments and the actual metered physical positions of each participant. 
This approach is predominant in Europe and widespread elsewhere. Within this context, the rapid expansion of renewable resources and demand-side engagement has introduced substantial uncertainties that manifest close to delivery, and this has translated into serious settlement risks for the market participants~\cite{SCHARFF2016544}. 
As a consequence, the risk management needs of participants have been met, partially at least, by the emergence of continuous intraday (CID) electricity trading to facilitate the adjustment of physical positions as new information
becomes apparent closer to delivery~\cite{koch2019short}.

\begin{figure}[!t]
  \hspace*{-2.2mm}
\includegraphics[width=1.03\linewidth]{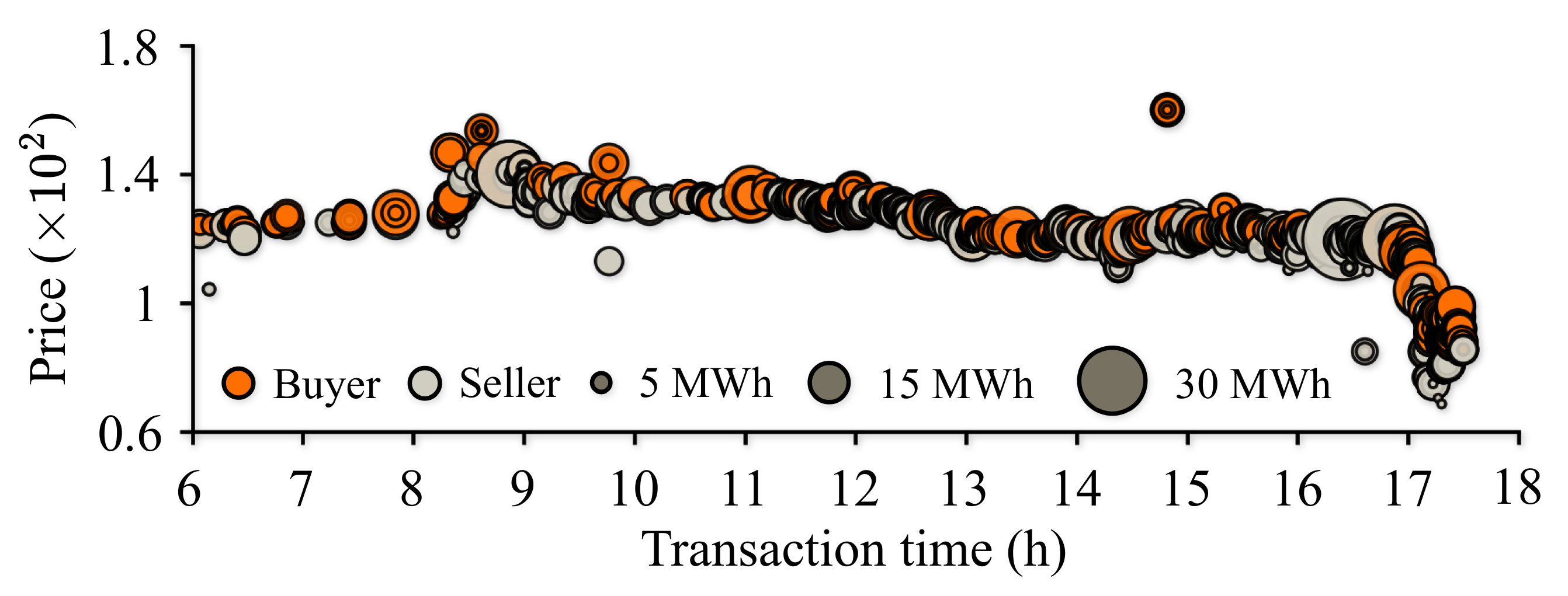}
\caption{
Buy-sell interactions for delivery at 18:00 on 2024-07-23. 
Buyers and sellers adjust their bids and offers iteratively based on the opposite side,
reflecting strategic buy-sell interactions. 
As delivery time approaches, prices exhibit downward jumps.
}
    \label{fig:buy-sell*}
\end{figure} 
In contrast to the auction-based day-ahead markets, CID trading is typically based upon bids and offers being continuously posted on a trading platform and, as such, presents a fundamentally different challenge for time-series analysis and forecasting compared to the more regular, distinct, and episodic price series usually obtained from auction data. As a research topic, CID electricity price forecasting has so far received much less attention than, for example, day-ahead price forecasting~\cite{LEHNA2022105742}. Furthermore, being motivated by the risk management needs of asset-backed traders, as well as perhaps by purely financial speculative traders, probabilistic forecasts will generally be more useful than point estimates. This compounds the modeling challenge. Whilst there has been extensive research on probabilistic day-ahead forecasting~\cite{9788043, variance_stab, yu2026pricefmfoundationmodelprobabilistic, 11045523, 10012043}, and an emerging body of research on the properties of the intraday price dynamics~\cite{drek, 24Multivariate, nickelsen2025bayesian, yu2026deeplearningelectricityprice},  a fully informative methodology for probabilistic CID forecasting remains under-researched. 

\begin{figure*}[!t]
\begin{center}
    \includegraphics[width=1\linewidth]{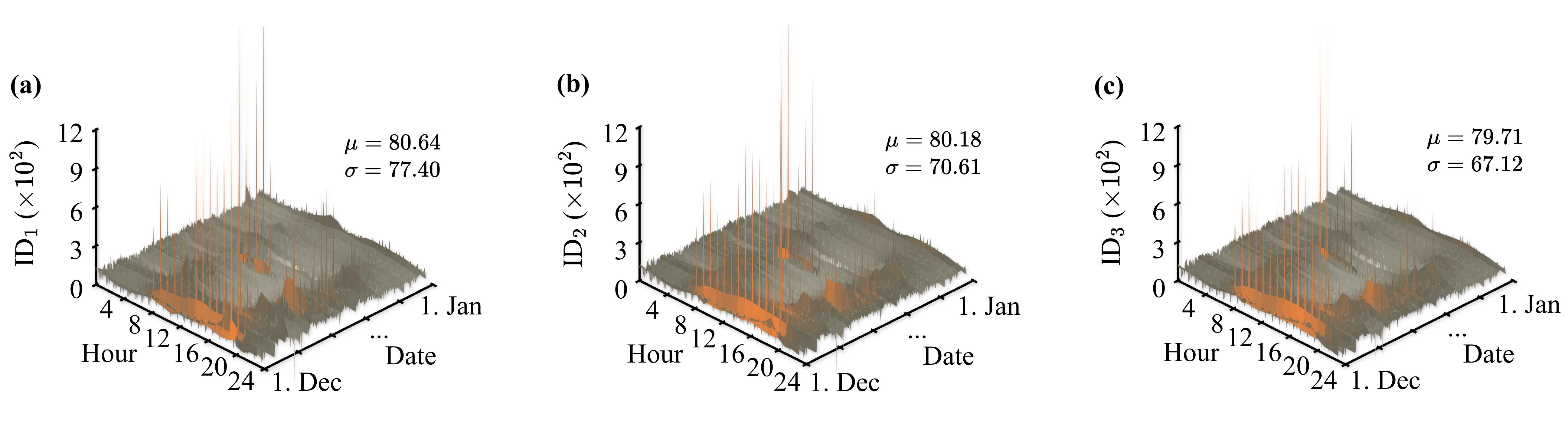}
\caption{
Visualization of CID price indices in the German market (2024). Orange areas indicate extreme electricity price. 
Overall, volatility follows the order: ID$_1$ $>$ ID$_2$ $>$ ID$_3$. 
\textbf{(a)}, ID$_1$ displays frequent price spikes, reflecting last-minute trading under imbalance pressure. 
\textbf{(b)}, ID$_2$ reflects mid-session adjustments. 
\textbf{(c)}, ID$_3$ corresponds to the most liquid trading window, exhibiting the least volatility. 
}
\label{orderbookvisualization*}
\end{center}
\end{figure*}

Moreover, whilst continuous trading is the norm in financial capital markets, intraday electricity trading is fundamentally different than typical financial orderbook settings: as an energy commodity, traders in the CID electricity market submit bids and offers for electricity tied to specific delivery times~\cite{Ensemble}.
The first problem to be faced in practice, therefore, is to define the reference price to be forecast ahead of delivery. 
A common simple heuristic is to take the average of the bid and offer prices, but as an index, this will generally be too volatile and subject to price jumps close to electricity delivery times~\cite{price_jumps}. 
\textcolor{black}{To support liquidity in trading and to offer more stable reference prices, the European power exchange (EPEX Spot) provides CID indices: \( \mathrm{ID}_1 \), \( \mathrm{ID}_2 \), and \( \mathrm{ID}_3 \), which are defined as Volume-Weighted Average Prices (VWAPs) of executed bids and offers aggregated over progressively longer  periods prior to delivery, as illustrated in Fig.~\ref{orderbookvisualization*}.}

\textcolor{black}{CID prices are not merely financial signals: they provide operationally relevant information for imbalance-risk management and demand-side flexibility. Accurate probabilistic forecasts of these prices can therefore support uncertainty-aware decision-making by balancing-responsible parties and asset operators under increasing renewable variability. This places probabilistic CID price forecasting within the broader scope of power system operation, computing, and economics.}

Conventional methods for forecasting these CID indices rely heavily on 1D encoding via domain feature extraction (e.g., mean price, total traded volume, and VWAP)~\cite{Trading, Simulation, yu2025orderbookfeaturelearningasymmetric}, directly aggregating over buy and sell sides. 
This aggregation overlooks buy-sell interactions~\cite{nagel1995unraveling}. 
The close buy-sell relationship is shown in Fig. \ref{fig:buy-sell*}. Moreover, this manual feature extraction breaks the end-to-end learning, restricting the model from forming expressive representations from the orderbook. 
In contrast to 1D encoding, advanced time-series models encode the orderbook as 2D sequence~\cite{FEDFormer, iTrans, PatchTST, TimesNet, TimeXer}, preserving the temporal information. 
However, these models still lack the inductive bias arising from buy–sell interactions, requiring large number of parameters to approximate such dynamics and leading to suboptimal performance. 
Motivated by the unique structure of the CID electricity market, we propose {OrderFusion}, which is designed for delivery-tied electricity trading and explicitly captures buy-sell interaction-aware representation from the orderbook data.



Furthermore, when modeling probabilistic intraday prices, a common approach to avoid a specific functional form for the density, is Linear Quantile Regression (LQR), in which each quantile is modeled separately~\cite{sortquantilecrossing1, sortquantilecrossing2, Trading}. 
However, this can introduce the quantile crossing problem, whereby higher quantile forecasts occasionally fall below lower ones, violating the monotonicity of the predictive distribution. We introduce constraints into our methodology to avoid this problem.
In summary, the contributions of this paper are:
\begin{itemize}
    \item \textcolor{black}{We formulate probabilistic CID price-index forecasting as a microstructure-aware power-system economics problem, motivated by the operational need for uncertainty-aware position adjustment close to delivery.}
    \item We propose OrderFusion, an end-to-end and parameter-efficient forecasting model that explicitly encodes buy-sell interactions in CID orderbooks while producing non-crossing probabilistic forecasts.
    \item We conduct exhaustive experiments to compare OrderFusion against multiple baselines and examine its generalizability across markets with high and low liquidity.
    \item We perform ablation studies to assess the impact of each architectural design, revealing the contribution of each component to overall predictive performance.
\end{itemize}

    


\section{Related Work}
\textbf{1D Encoding.}
A commonly used domain feature is the VWAP over the past 15 minutes ~\cite{beating, Econometric, Trading}, which has shown strong empirical performance. 
Other studies indicate that the last price already reflects past information, implying weak-form efficiency\footnote{Under the Efficient Market Hypothesis (EMH), a market is weak-form efficient if recent prices already reflect predictive information contained in historical orders~\cite{fama1970efficient}.} \cite{ Forecasting, Understanding, Simulation}.
Moreover, \cite{yu2025orderbookfeaturelearningasymmetric} extracts \textcolor{black}{numerous} domain features from the orderbook and reveals that price percentiles are strong predictors. 
However, relying solely on manual feature extraction neglects the inductive biases arising from buy–sell interactions, degrading predictive performance.

\textbf{2D Encoding.}
Advanced time-series models such as FEDFormer~\cite{FEDFormer}, iTransformer~\cite{iTrans}, PatchTST~\cite{PatchTST}, TimesNet~\cite{TimesNet}, and TimeXer~\cite{TimeXer} represent the orderbook as a 2D multivariate sequence and have achieved notable success in various forecasting tasks by capturing complex temporal patterns. 
However, these models do not explicitly encode the inductive bias of buy-sell interactions, and thus require substantially more parameters to approximate such dynamics, resulting in suboptimal performance.

\textbf{Quantile Crossing.}
When multiple conditional quantiles are estimated independently, quantile crossing may occur. 
Non-crossing can be enforced using monotone networks via derivative integration formulations or gradient penalties~\cite{brando2022deep, monteiro2022monotonicity}. 
However, these approaches require careful tuning of polynomial degree or penalty strength, increasing training complexity. 
A simple and widely used method in the field of electricity price forecasting is post-hoc sorting, which reorders the predicted quantiles~\cite{sortquantilecrossing1, sortquantilecrossing2, Trading}. 
While effective in practice, such method introduces an additional correction step in the forecasting pipeline, breaking the end-to-end learning.

\section{Preliminary}
\label{sec:preliminary}
The task is to forecast three widely used price indices 
\textcolor{black}{in the European CID market}: 
\( \mathrm{ID}_x \), where \( x \in \{1, 2, 3\} \), 
\textcolor{black}{which differ by the length of the trading window}. 
\textcolor{black}{Formally}, each \( \mathrm{ID}_x \) is defined as the VWAP of trades executed over its corresponding trading window:

\begin{equation}
\mathrm{ID}_x
\;=\;
\frac{\sum\limits_{s \in S} \sum\limits_{t \in \mathcal{T}_{f}} P^{(s)}_{t} \, V^{(s)}_{t}}
     {\sum\limits_{s \in S} \sum\limits_{t \in \mathcal{T}_{f}} V^{(s)}_{t}},
\end{equation}
where the forecast is made at time \( t_f = t_d - \Delta \), with \( t_d \) denoting the delivery time and 
\( \Delta = 60 \times x\,\mathrm{min} \) representing the lead time associated with index \( \mathrm{ID}_x \).
The market side \( s \in S = \{+, -\} \) corresponds to buy and sell orders, respectively, 
\( t \in \mathcal{T}_{f} = [t_f,\ t_d - \delta_c] \) denotes the transaction time, 
\( \mathcal{T}_{f} \) is the forecasting (trading) window, and 
\( \delta_c \) is a market-specific parameter\footnote{For Germany, \( \delta_c = 30 \) min, 
and for Austria, \( \delta_c = 0 \) min. For other countries, \( \delta_c \) can be retrieved from 
\textcolor{orange}{\url{https://www.epexspot.com/en/downloads}}.}.
Here, \( P^{(s)}_{t} \) and \( V^{(s)}_{t} \) denote the price and traded volume, respectively. 



\section{OrderFusion}
\label{sec:OrderFusion}

\begin{figure*}[!t]
\centering
\hspace*{-2.2mm}
    \includegraphics[width=1.02\linewidth]{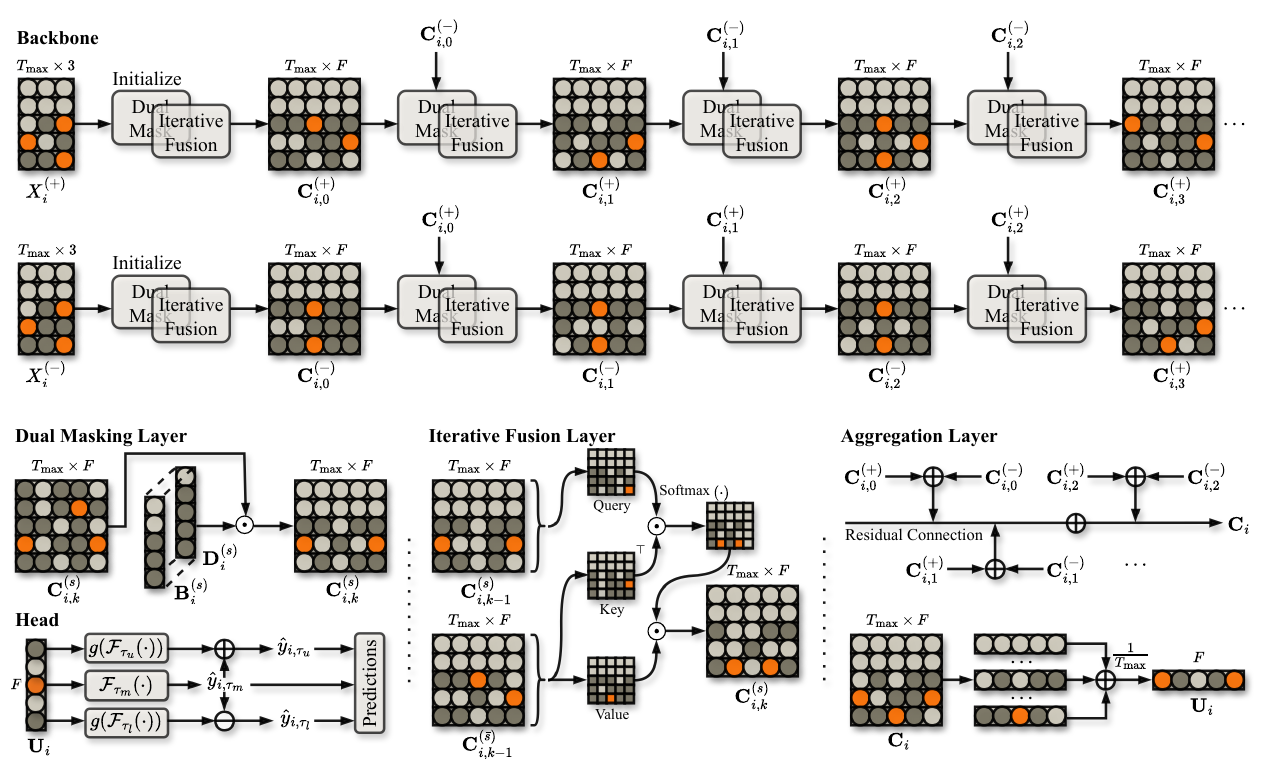}
\caption{
Structure of OrderFusion. 
The buy-side input and sell-side input are masked through dual masking layers and iteratively fused to form representations of buy–sell interactions in the latent space. The representations are aggregated across different degrees of interactions and then passed through a hierarchical head to generate multiple quantile estimates, enabling end-to-end probabilistic forecasting.}
\label{OrderFusion_Structure}
\end{figure*}

As introduced in Section~\ref{introduction}, the CID market is driven by the interaction between buy-side and sell-side orders. Therefore, such buy-sell interactions should be explicitly modeled. 
To this end, OrderFusion is designed to exploit the predictive potential of orderbook data (Fig.~\ref{OrderFusion_Structure}). Specifically,
we introduce an encoding method in 
Subsection~\ref{sec:OrderFusionEncodingMethod} to transform the orderbook into an encoded format that can be used by the backbone.
The backbone described in Subsection~\ref{sec:backbone} processes the encoded orderbook and produces buy-sell-interaction-aware representations through a cross-attention mechanism. 
The forecasting head in Subsection~\ref{sec:hierchical_head} produces multiple quantile forecasts while preventing quantile crossing.

\subsection{Encoding}
\label{sec:OrderFusionEncodingMethod}

The encoding method separates the orderbook into two sides: buy (\(+\)) and sell (\(-\)). For each side, we treat all trades associated with each delivery time as one sample. 
Each trade contains a price and a traded volume. Additionally, we compute a relative time delta \( \nabla t \) to encode temporal information:
\begin{equation}
    \nabla t = t_d - t, \quad  t < t_f.
\end{equation}

Notably, the number of trades varies across samples, as trades are irregularly distributed over transaction time. Therefore, each sample is represented as a variable-length 2D sequence:
\begin{equation}
    X^{(s)}_i = 
    \begin{bmatrix}
        P^{(s)}_{t_1} & V^{(s)}_{t_1} & \nabla t_1 \\
        P^{(s)}_{t_2} & V^{(s)}_{t_2} & \nabla t_2 \\
        \vdots        & \vdots        & \vdots       \\
        P^{(s)}_{t_j} & V^{(s)}_{t_j} & \nabla t_j \\
        \vdots        & \vdots        & \vdots       \\
        P^{(s)}_{t_{T_i^{(s)}}} & V^{(s)}_{t_{T_i^{(s)}}} & \nabla t_{T_i^{(s)}}
    \end{bmatrix}
\end{equation}
where \( X^{(s)}_i \in \mathbb{R}^{T_i^{(s)} \times 3}\) is the input matrix for the \( i \)-th sample on side \( s \), with \( T_i^{(s)} \) denoting the number of trades. The index \( i \in \{1, 2, \dots, N\} \) enumerates samples, and \( j \in \{1, 2, \dots, T_i^{(s)}\} \) denotes the \( j \)-th timestep within sample \( i \). 

The encoded data consists of two time series matrices, one for each side of the orderbook (buy, sell) in which the time series of orders is related to feature time series of technical indices. This {2$\times$2D}
 representation therefore consists of two irregular 2D sequences per sample for the buy and sell sides:
\begin{equation}
    \mathcal{X}^{(+)} = \left\{ X^{(+)}_1, \dots, X^{(+)}_i, \dots, X^{(+)}_N \right\}, 
\end{equation}
\begin{equation}
    \mathcal{X}^{(-)} = \left\{ X^{(-)}_1, \dots, X^{(-)}_i, \dots, X^{(-)}_N \right\}.
\end{equation}
\subsection{Backbone}
\label{sec:backbone}
\subsubsection{Dual Masking Layer}  
As the number of matched trades varies between samples, we apply pre-padding to align all input sequences to a maximum length \( T_{\max} \). Padding values are set to a constant \( c = 10{,}000 \) to ensure they do not occur in the data. Thus, the input dimension is standardized to \( \mathbb{R}^{T_{\max} \times 3} \). To identify valid timesteps, we define a binary padding mask \( \mathbf{B}^{(s)}_i \in \{0, 1\}^{T_{\max} \times 1} \) as:
\begin{equation}
    \mathbf{B}^{(s)}_i[j] = 
    \begin{cases}
        1 & \text{if } {X}^{(s)}_i[j, :] \neq c, \\
        0 & \text{otherwise.}
    \end{cases}
\end{equation}
To reflect the prior that recent trades carry the most predictive information under the market efficiency hypothesis~\cite{hasbrouck1991measuring, hasbrouck2007empirical, bacry2015hawkes}, 
we define a binary temporal mask \( \mathbf{D}^{(s)}_i \in \{0, 1\}^{T_{\max} \times 1} \), where the cutoff length is given by \( L = 2^\alpha \), controlled by a hyperparameter \( \alpha \in \mathbb{N} \), with \( L \leq T_{\max} \):
\begin{equation}
    \mathbf{D}^{(s)}_i[j] = 
    \begin{cases}
        1 & \text{if } j > T_{\max} - L, \\
        0 & \text{otherwise.}
    \end{cases}
\label{cutoff_eq}
\end{equation}
The dual mask is obtained by elementwise multiplication of the padding and temporal masks:
\begin{equation}
    \mathbf{M}_{i}^{(s)} = \mathbf{B}_{i}^{(s)} \odot \mathbf{D}_{i}^{(s)}.
    \label{eq:dualmask}
\end{equation}
\subsubsection{Iterative Fusion Layer}
As buyers and sellers iteratively adjust their bids and offers based on observed quotes from the opposite side, reflecting strategic interactions~\cite{nagel1995unraveling}, we design a series of iterative fusion layers to enable representation learning of such buy-sell interactions:
\begin{equation}
    \mathbf{C}_{i,k}^{(s)} = 
    \begin{cases}
        \mathcal{F}^{(s)}\!\left(X^{(s)}_i\right)  & \text{if } k = 0, \\
        \mathbf{C}_{i,k-1}^{(s)} \mid \mathbf{C}_{i,k-1}^{(\bar{s})} & \text{if } k \ge 1,
    \end{cases}
    \label{eq:jumpfusion}
\end{equation}
where \( k \) denotes the degree of interactions, \( \bar{s} \) is the opposite side of \( s \), and \(\mathcal{F}^{(s)}(\cdot)\) denotes a dense layer that projects the input features from dimension \(3\) to the hidden dimension \(F\). All intermediate representations are masked using \( \mathbf{M}_{i}^{(s)} \) before being passed to subsequent layers.

For \( k = 0 \), the representation is initialized by projecting the masked input matrix \(X^{(s)}_i\) into the hidden dimension, such that \(\mathbf{C}_{i,0}^{(s)} \in \mathbb{R}^{T_{\max}\times F}\).

For \( k \ge 1 \), a cross-attention is applied, denoted by the fusion operator ``\( \mid \)'', where \( \mathbf{C}_{i,k-1}^{(s)} \) serves as the query, and \( \mathbf{C}_{i,k-1}^{(\bar{s})} \) serves as both the key and value:
\begin{equation}
\mathbf{Q}_{k-1}^{(s)} = \mathbf{C}_{i,k-1}^{(s)} \mathbf{W}_{\mathbf{Q}, k-1}^{(s)}, 
\label{eq:q_mainbody}
\end{equation}
\begin{equation} 
\mathbf{K}_{k-1}^{(\bar{s})} = \mathbf{C}_{i,k-1}^{(\bar{s})} \mathbf{W}_{\mathbf{K}, k-1}^{(\bar{s})}, 
\label{eq:k_mainbody}
\end{equation}
\begin{equation}
\mathbf{V}_{k-1}^{(\bar{s})} = \mathbf{C}_{i,k-1}^{(\bar{s})} \mathbf{W}_{\mathbf{V}, k-1}^{(\bar{s})}.
\label{eq:v_mainbody}
\end{equation}
where \( \mathbf{W}_{\mathbf{Q}, k-1}^{(s)}, \mathbf{W}_{\mathbf{K}, k-1}^{(\bar{s})}, \mathbf{W}_{\mathbf{V}, k-1}^{(\bar{s})} \in \mathbb{R}^{F \times F} \) are learnable weights, and \( F \) denotes the hidden dimension. The output of the cross-attention is computed as:
\begin{equation}
\mathbf{C}_{i,k}^{(s)} = \mathrm{Softmax} 
\left( \frac{\mathbf{Q}_{k-1}^{(s)} (\mathbf{K}_{k-1}^{(\bar{s})})^\top}{\sqrt{F}} \right) \mathbf{V}_{k-1}^{(\bar{s})}.
\label{eq:crossattn_output_mainbody}
\end{equation}
This allows side \( s \) to observe the opposite side and form updated representations that reflect buy–sell interactions. For example, \( \mathbf{C}_{i,1}^{(+)} \) denotes for the buy-side representation observed on sell-side. 

\subsubsection{Aggregation Layer}
All of the fused representations 
at different degrees are combined via residual connection \cite{xie2017aggregated} to produce the higher-level representation \( \mathbf{C}_{i} \in \mathbb{R}^{T_{\max} \times F} \):
\begin{equation}
    \mathbf{C}_i = \sum_{k=1}^{K} \left( \mathbf{C}_{i,k}^{(+)} + \mathbf{C}_{i,k}^{(-)} \right),
    \label{eq:ResRail}
\end{equation}
where \( K \) denotes the maximum degree of interactions, and the summation is element-wise addition.

We apply global average pooling to obtain the attention-weighted average representation \( \mathbf{U}_{i} \in \mathbb{R}^{F} \):
\begin{equation}
    \mathbf{U}_{i} = \frac{1}{T_{\max}} \sum_{j=1}^{T_{\max}} \mathbf{C}_i[j].
    \label{eq:avgpool}
\end{equation}
\subsection{Head}
\label{sec:hierchical_head}
The hierarchical head produces multiple quantile forecasts, where \( \tau \in \mathcal{Q} = \{0.10, 0.25, 0.45, 0.50, 0.55, 0.75, 0.90\} \).
In detail, the median quantile (\( \tau_m = 0.50 \)) is directly predicted through a dense layer \( \mathcal{F}_{\tau_m}(\cdot) \):
\begin{equation}
    \hat{y}_{i, \tau_m} = \mathcal{F}_{\tau_m}(\mathbf{U}_{i}).
    \label{eq:median}
\end{equation}

For each upper quantile \( \tau_u > 0.50 \), a residual is generated from \( \mathbf{U}_{i} \) via a separate dense layer \( \mathcal{F}_{\tau_u}(\cdot) \).  
The residual is enforced to be non-negative by applying an absolute-value function \( g(\cdot) = |\cdot| \).
The upper quantile is then computed hierarchically by adding this non-negative residual to its nearest lower quantile \(\tau_{u'}\):
\begin{equation}
    \hat{y}_{i, \tau_u} = 
    \hat{y}_{i, \tau_{u'}} + g\!\left(\mathcal{F}_{\tau_u}(\mathbf{U}_{i})\right),
    \quad \forall\, \tau_u > \tau_m.
    \label{eq:upper_hier}
\end{equation}

For each lower quantile \( \tau_l < 0.50 \), a non-negative residual is similarly generated and subtracted from its nearest higher quantile:
\begin{equation}
    \hat{y}_{i, \tau_l} = 
    \hat{y}_{i, \tau_{l'}} - g\!\left(\mathcal{F}_{\tau_l}(\mathbf{U}_{i})\right),
    \quad \forall\, \tau_l < \tau_m.
    \label{eq:lower_hier}
\end{equation}

\subsection{Loss}
\label{loss}
Average Quantile Loss (AQL) is employed to jointly estimate multiple quantiles:
\begin{equation}
    \text{AQL} = \frac{1}{ N|\mathcal{Q}|} \sum_{i=1}^N \sum_{\tau \in \mathcal{Q}}  L_\tau(y_i,  \hat{y}_{i, \tau}),
    \label{AQL_loss}
\end{equation}
where \( y_i \) is the true price, \( \hat{y}_i \) denotes the predicted price quantile, and the loss \( L_\tau \) is defined as:
\begin{equation}
    L_\tau(y_i, \hat{y}_{i, \tau}) = 
    \begin{cases} 
      \tau \cdot (y_i -  \hat{y}_{i, \tau}), & \text{if } y_i \geq  \hat{y}_{i, \tau}, \\
      (1 - \tau) \cdot ( \hat{y}_{i, \tau} - y_i), & \text{otherwise}.
    \end{cases}
\end{equation}

\section{Baselines}
\subsection{Na\"ive} 
    \textbf{Na\"ive$^1$:} The price index from the most recent delivery hour is used as the na\"ive point forecast, which is generally a strong and hard-to-beat baseline. 
    \textbf{Na\"ive$^2$:} The price index from the same delivery hour on the previous day is used. \textbf{Na\"ive$^3$:} The price index from the same delivery hour, averaged over the past 3 days, is used.
To obtain probabilistic forecasts, we first compute the residuals as the difference between the true price and the na\"ive point forecast from the training data. Then, these residuals are grouped by delivery hour to estimate their percentiles. Lastly, we add these hour-specific residual percentiles to the na\"ive point forecasts to form the probabilistic forecasts.

\subsection{1D Encoding}  

The 1D encoding compresses the orderbook into domain features. 
We use three representative feature baselines: the \textbf{15-min VWAP}, the \textbf{last price}, and the \textbf{exhaustive feature set (384 features)}. We evaluate both a non-linear model (MLP) and a linear model (LQR) for each feature baseline, reporting only the better results to avoid model-specific bias. 

    \textbf{15-Min VWAP.} Prior studies report that the VWAP over the last 15 minutes is a strong domain feature for both pointwise and probabilistic forecasting~\cite{beating, Econometric, Trading}. We compute it as:
    \begin{equation}
\text{VWAP}\big|_{\mathcal{T}_{15}}
= \frac{\sum\limits_{t \in \mathcal{T}_{15}} P^{(s)}_{t} \, V^{(s)}_{t}}
     { \sum\limits_{t \in \mathcal{T}_{15}} V^{(s)}_{t}}
\end{equation}

    where \(
\mathcal{T}_{15} = [t_f - 15, \, t_f]
\) denotes the look-back window for feature extraction from the last 15 minutes.

   \textbf{Last Price.} Existing studies indicate that the last price already reflects past information~\cite{Short, Probab, Forecasting, Understanding, Simulation}, implying weak-form efficiency. We use this baseline to examine whether the CID market exhibits perfect weak-form efficiency. We compute it as:
    \begin{equation}
\text{LastP}\big|_{\mathcal{T}_\infty} = P^{(s)}_{t_\text{max}},
\end{equation}
where \(
\mathcal{T}_{\infty} = [t_f - \infty,\, t_f]
\) denotes the full look-back window starting from market opening.
    
    \textbf{Exhaustive Feature Set.} An extensive set of features, such as min, max, and volatility of prices and traded volumes, is extracted from multiple look-back windows 
\(
\mathcal{T}_{w} = [t_f - \delta_w, \, t_f]
\), where \( \delta_w \in \{1, 5, 15, 60, 180, \infty\} \) (in minutes), totaling 384 features~\cite{yu2025orderbookfeaturelearningasymmetric}. Feature selection is conducted using \( \ell_1 \)-regularization.

\subsection{2D Encoding}
The 2D encoding preserves the temporal dimension of the orderbook by representing it as a multivariate time series. 
To assess the predictive capability of 2D encoding, we benchmark five state-of-the-art time-series models: \textbf{FEDFormer}, \textbf{iTransformer}, \textbf{PatchTST}, \textbf{TimesNet}, \textbf{TimeXer}. As these model do not support a {2$\times$2D} input, we concatenate the buy-side and sell-side sequences along the feature dimension to form a multivariate time series.
The hyperparameter is optimized based on validation performance via exhaustive grid search.

\textbf{FEDFormer.} The frequency-enhanced decomposition and Fourier attention are introduced to capture periodic patterns~\cite{FEDFormer}. 
The model could be useful if intraday prices exhibit strong seasonal periodicity.

\textbf{iTransformer.} The input variables are treated as tokens, reducing attention dimensionality and enabling efficient feature learning~\cite{iTrans}. 
The model could be beneficial for intraday price forecasting when long sequences are used as input.

\textbf{PatchTST.} The time series is patchified, and a channel-independent Transformer is used to improve the representation of local temporal patterns~\cite{PatchTST}. 
The model is potentially useful if local price fluctuations dominate predictive performance.

\textbf{TimesNet.} The multi-scale kernels are learned through  convolutions in the frequency domain~\cite{TimesNet}. 
The model could be valuable if prices exhibit different temporal patterns.

\textbf{TimeXer.} The temporal compression is combined with cross-scale mixing to model both fine-grained and aggregated temporal representations~\cite{TimeXer}. 
The model could be beneficial for capturing rapid price jumps and broader market trends.


 


\begin{table}[t]
\centering
\caption{Model comparison for the German market.}
\label{performance_compare_table_DE_compact}
\begin{tabular}{
>{\raggedright\arraybackslash}p{1.55cm}
>{\centering\arraybackslash}p{0.9cm}
>{\centering\arraybackslash}p{1.1cm}
>{\centering\arraybackslash}p{0.9cm}
>{\centering\arraybackslash}p{1.1cm}
>{\centering\arraybackslash}p{0.9cm}
}
\toprule
\multirow{2}{*}{\textbf{Model}}
& \multicolumn{2}{c}{\cellcolor{customorange!80}\textbf{Probabilistic}}
& \multicolumn{2}{c}{\cellcolor{customorange!80}\textbf{Pointwise}}
& \multirow{2}{*}{\textbf{Rank}} \\
\cmidrule(lr){2-3}
\cmidrule(lr){4-5}
& \textbf{AQL} $\downarrow$
& \textbf{AQCR} $\downarrow$
& \textbf{MAE} $\downarrow$
& \textbf{RMSE} $\downarrow$
& \\
\midrule

Na\"ive$^{1}$
& 6.34 & \textcolor{gray}{0.00} & 14.75 & 40.11 & 9 \\

Na\"ive$^{2}$
& 16.17 & \textcolor{gray}{0.00} & 36.62 & 85.66 & 11 \\

Na\"ive$^{3}$
& 15.83 & \textcolor{gray}{0.00} & 36.62 & 76.63 & 10 \\

\midrule

MLP $\vert$ LQR$^{1}$
& 4.72 & 0.19 & 11.43 & 34.21 & 6 \\

MLP $\vert$ LQR$^{2}$
& 4.87 & 0.46 & 11.89 & 45.46 & 8 \\

MLP $\vert$ LQR$^{3}$
& 4.61 & 0.54 & 10.96 & 32.23 & 3 \\

\midrule

FEDFormer
& 4.70 & 3.96 & 11.52 & 34.49 & 8 \\

iTransformer
& 4.66 & 3.77 & 11.55 & 33.99 & 7 \\

PatchTST
& 4.40 & 2.99 & 11.27 & 32.34 & 4 \\

\rowcolor{DarkOrange!10}
TimesNet
& 4.38 & 2.57 & 10.98 & 32.20 & 2 \\

TimeXer
& 4.53 & 2.87 & 11.32 & 33.01 & 5 \\

\midrule

\rowcolor{customgray!20}
\textbf{OrderFusion}
& \textbf{3.81}
& \textbf{0.00}
& \textbf{9.06}
& \textbf{26.84}
& \textbf{1} \\

\bottomrule
\end{tabular}
\end{table}

\section{Experiments}
\label{sec:exp}

\subsection{Experimental Settings}

\textbf{Rolling Evaluation.}
We employ a 3-fold rolling evaluation. In fold~1, the data span from 1.~Jan~2022 to 1.~Sep~2023 for training, 1.~Sep~2023 to 1.~Jan~2024 for validation, 
and 1.~Jan~2024 to 1.~May~2024 for testing. 
Each subsequent fold shifts forward by 4 months until reaching 1.~Jan~2025 to ensure that the testing periods collectively span a full year~\cite{LAGO2021116983}.

\textbf{Data Scaling.} We use the \texttt{RobustScaler} to scale the data for being robust to extreme values. The scaler is fitted on the training data, and the fitted scaler is used to transform validation and testing data. 

\textbf{Evaluation Metrics.}
We utilize AQL and Average Quantile Crossing Rate (AQCR) to evaluate the probabilistic forecasting accuracy.
The Mean Absolute Error (MAE) and Root Mean Squared Error (RMSE) are used for pointwise evaluation. 

\textbf{Loss Aggregation.}
The testing losses are aggregated from 3 rolling folds, 3 price indices, and 5 random seeds for the German and Austrian markets, respectively, to ensure a robust evaluation. 
\textcolor{black}{The German and Austrian CID markets are selected to evaluate the proposed method under different liquidity regimes~\cite{yu2025orderbookfeaturelearningasymmetric}. 
From a power-system operation perspective, market liquidity is important as it determines how rapidly renewable forecast updates and balancing needs are reflected in intraday prices.}







\subsection{Empirical Results}
The main experimental results are summarized in Table~\ref{performance_compare_table_DE_compact} and Table~\ref{performance_compare_table_AT_compact}. 
Across both the German and Austrian markets, OrderFusion consistently achieves superior probabilistic and pointwise forecasting accuracy.
The detailed observations are summarized as follows.




\begin{table}[t]
\centering
\caption{Model comparison for the Austrian market. }
\label{performance_compare_table_AT_compact}
\begin{tabular}{
>{\raggedright\arraybackslash}p{1.55cm}
>{\centering\arraybackslash}p{0.9cm}
>{\centering\arraybackslash}p{1.1cm}
>{\centering\arraybackslash}p{0.9cm}
>{\centering\arraybackslash}p{1.1cm}
>{\centering\arraybackslash}p{0.9cm}
}
\toprule
\multirow{2}{*}{\textbf{Model}}
& \multicolumn{2}{c}{\cellcolor{customorange!80}\textbf{Probabilistic}}
& \multicolumn{2}{c}{\cellcolor{customorange!80}\textbf{Pointwise}}
& \multirow{2}{*}{\textbf{Rank}} \\
\cmidrule(lr){2-3}
\cmidrule(lr){4-5}
& \textbf{AQL} $\downarrow$
& \textbf{AQCR} $\downarrow$
& \textbf{MAE} $\downarrow$
& \textbf{RMSE} $\downarrow$
& \\
\midrule

Na\"ive$^{1}$
& 8.46 & \textcolor{gray}{0.00} & 19.96 & 42.95 & 8 \\

Na\"ive$^{2}$
& 15.62 & \textcolor{gray}{0.00} & 36.59 & 72.09 & 10 \\

Na\"ive$^{3}$
& 15.25 & \textcolor{gray}{0.00} & 36.00 & 64.95 & 9 \\

\midrule

MLP $\vert$ LQR$^{1}$
& 6.58 & 0.15 & 15.64 & 34.39 & 5 \\

MLP $\vert$ LQR$^{2}$
& 6.67 & 0.24 & 15.84 & 37.38 & 6 \\

MLP $\vert$ LQR$^{3}$
& 6.49 & 0.44 & 15.12 & 33.65 & 2 \\

\midrule

FEDFormer
& 6.52 & 4.97 & 16.35 & 34.47 & 7 \\

iTransformer
& 6.44 & 3.87 & 16.41 & 34.15 & 6 \\

PatchTST
& 6.19 & 2.29 & 15.17 & 33.94 & 3 \\

\rowcolor{DarkOrange!10}
TimesNet
& 6.15 & 2.32 & 15.14 & 33.70 & 2 \\

TimeXer
& 6.24 & 3.05 & 15.22 & 33.88 & 4 \\

\midrule

\rowcolor{customgray!20}
\textbf{OrderFusion}
& \textbf{5.64}
& \textbf{0.00}
& \textbf{13.52}
& \textbf{29.95}
& \textbf{1} \\

\bottomrule
\end{tabular}
\end{table}
\textbf{Against Na\"ive Models.}
The Na\"ive baselines are consistently inferior to OrderFusion across both probabilistic and pointwise evaluation. 
The AQL averaged across price indices and markets of Na\"ive$^1$ is \textbf{56.69\%} higher than that of OrderFusion, while Na\"ive$^2$ and Na\"ive$^3$ are \textbf{236.40\%} and \textbf{228.90\%} higher, respectively. 
A similar pattern is observed for pointwise forecasting.
This suggests that Na\"ive baselines are insufficient for intraday price prediction, as it cannot capture the nonlinear and dynamic interactions between buy-side and sell-side order flows. 
Therefore, the buy-sell interaction needs to be explicitly modeled rather than relying solely on simple persistence.

\textbf{Against 1D Encoding Models.}
OrderFusion achieves AQL improvements of \textbf{16.37\%}, \textbf{18.13\%}, and \textbf{14.86\%} over the 15-min VWAP, last-price, and exhaustive-feature baselines, respectively. 
This indicates that domain-feature-based methods still underutilize the predictive information contained in the raw orderbook. 
In particular, the improvement over the last-price baseline suggests that the CID market is not perfectly weak-form efficient, as historical buy-sell order flow continues to carry predictive information beyond the most recent transaction price. 
Moreover, these 1D-encoded methods suffer from quantile crossing, indicated by nonzero AQCR values. 
Although the observed violation rate is relatively low, this issue is expected to become more pronounced when predicting a larger number of quantiles~\cite{Trading}. 
By design, OrderFusion eliminates quantile crossing and consistently achieves an AQCR of \textbf{0.00\%}.

\textbf{Against 2D Encoding Models.} 
OrderFusion consistently outperforms 2D-encoding baselines and achieves \textbf{10.26\%} lower AQL and \textbf{14.09\%} lower MAE than the second-best performing TimesNet. 
This performance gap is attributed to the fact that these models are designed for generic time-series tasks and lack inductive biases for explicitly modeling buy-sell interactions. 
Moreover, these time-series baselines encounter significant quantile crossing issues, indicated by AQCR values between \textbf{1.97\%} and \textbf{4.97\%}. 
Furthermore, while these baselines contain between \textbf{0.87M} and \textbf{3.38M} parameters, OrderFusion remains lightweight with only \textbf{4,872} parameters. 
\textcolor{black}{Such parameter efficiency is especially relevant for power-system operational deployment, where probabilistic forecasts may be repeatedly generated for balancing-responsible portfolios or storage assets under limited computing resources.}
These results emphasize the importance of injecting domain priors into model design instead of relying solely on stacking model parameters.



\begin{figure*}[!t]
\hspace*{-1.9mm}
\centering
    \includegraphics[width=1.02\linewidth]
    {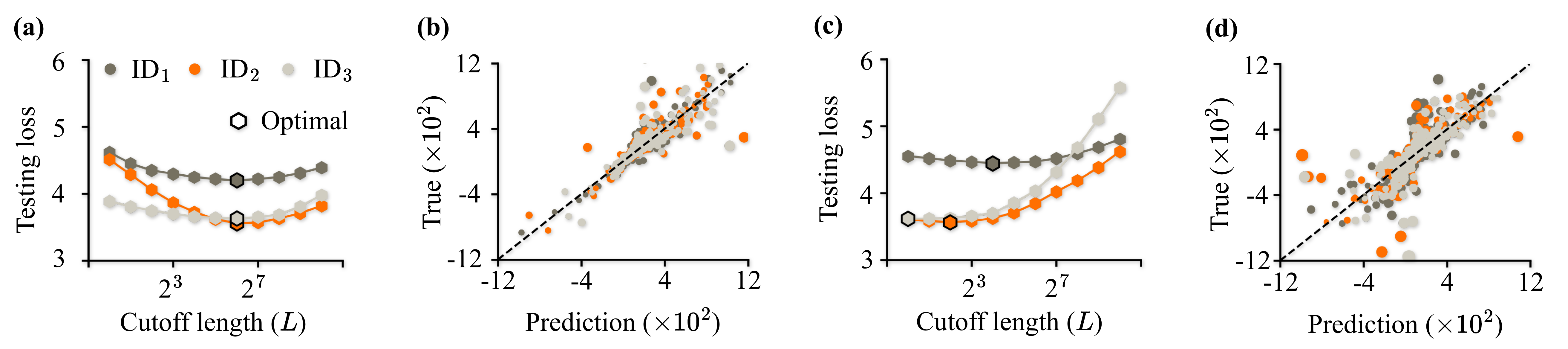}
\caption{
Ablation study of key hyperparameter (cutoff length) and empirical results of OrderFusion. 
\textbf{(a)} Testing loss under different cutoff lengths \(L\) for the German market, where the marked points indicate the optimal cutoff for each price index. 
\textbf{(b)} True price versus pointwise forecast for the German market. 
\textbf{(c)} Testing loss under different cutoff lengths \(L\) for the Austrian market. 
\textbf{(d)} True price versus pointwise forecast for the Austrian market. 
}
\label{fig:experiment}

\end{figure*}

\section{Ablation Study}
\label{sec:AblationStudy}

\subsection{Dual Masking Layer}
\begin{itemize}
    \item \textbf{No Mask.} Eq.~\ref{eq:dualmask} is removed, and no masks are applied in iterative fusion layers.

    \item \textbf{Random Mask.}  
Eq.~\ref{eq:dualmask} is replaced with a randomly sampled vector, where each element is independently drawn from a uniform distribution over \( [0, 1] \):
\begin{equation}
    \mathbf{M}_{i}^{(s)} \sim \mathcal{U}(0, 1),
    \label{eq:random_mask}
\end{equation}
    \item \textbf{Reverse Mask.} Instead of retaining the latest \( L \) buy and sell trades, we keep the first \( L \) buy and sell trades. Eq.~\ref{cutoff_eq} changes to:
\begin{equation}
    \mathbf{D}^{(s)}_i[j] = 
    \begin{cases}
        1 & \text{if } j \leq L, \\
        0 & \text{otherwise.}
    \end{cases}
\label{cutoff_reverse_eq}
\end{equation}

    \item \textbf{Varied Cutoff Length.} We vary the cutoff length \(L\) defined in Eq.~\ref{cutoff_eq} using \(\alpha = \{0,1,2,\ldots,10\}\) to examine the sensitivity of this key hyperparameter and to investigate market efficiency.
\end{itemize}

Results in Table~\ref{tab:ablation_study_compact} show that removing the mask leads to a \textbf{30.51}\% increase in AQL, as all padded values are treated as valid values, thereby introducing substantial noise. 
Randomizing and reversing the mask result in a \textbf{25.00\%} and \textbf{59.75\%} increase in AQL, respectively, emphasizing the importance of retaining only recent trades under the market efficiency hypothesis. Varying the cutoff length leads to significant changes in testing performance, as shown in Fig.~\ref{fig:experiment} \textbf{(a)} and \textbf{(b)}. 
These changes reflect the market efficiency. 
In particular, Austrian ID\(_3\)
has an optimal cutoff of 1, implying near-perfect market efficiency and indicating that traders rely on the latest pair of buy-sell trades to form price expectations.
Therefore, when applying OrderFusion to other markets, such as the Dutch or Norwegian markets, this cutoff hyperparameter should be tuned carefully. In practice, we recommend producing a similar testing-loss-versus-cutoff curve for each market and selecting the cutoff based on the observed performance.

\subsection{Iterative Fusion Layer}

\begin{itemize}
    \item \textbf{No Fusion.}  
     Eq.~\ref{eq:jumpfusion} is removed. The buy- and sell-side inputs are directly passed to subsequent layers without the representation learning of buy-sell interactions:
    \begin{equation}
        \mathbf{C}_{i} = X^{(+)}_i \, \| \, X^{(-)}_i,
        \label{eq:concat_fusion}
    \end{equation}
    where \( \big\| \) denotes concatenation.

    \item \textbf{Varied Degree of Interactions.} 
We vary the interaction degree \( K \) defined in Eq.~\ref{eq:jumpfusion}, using \( K \in \{1, 2, 4\} \), to study the necessary level of interactive complexity.


\end{itemize}

From Table~\ref{tab:ablation_study_compact}, we observe that removing the fusion layer results in an \textbf{18.86\%} increase in AQL, confirming that discarding the buy–sell inductive bias could degrade predictive performance. 
Moreover, the low-level interaction (\(K=1\)) leads to \textbf{2.54\%} higher AQL, while high-level interaction (\(K=4\)) shows no significant improvement over the medium setting (\(K=2\)), despite introducing additional parameters. Therefore, \(K=2\) is recommended for practical usage.


\begin{table}[t]
\centering
\caption{Ablation studies of different module choices.}
\label{tab:ablation_study_compact}
\begin{tabular}{
>{\raggedright\arraybackslash}p{1.55cm}
>{\centering\arraybackslash}p{0.9cm}
>{\centering\arraybackslash}p{1.1cm}
>{\centering\arraybackslash}p{0.9cm}
>{\centering\arraybackslash}p{1.1cm}
>{\centering\arraybackslash}p{0.9cm}
}
\toprule
\multirow{2}{*}{\textbf{Model}}
& \multicolumn{2}{c}{\cellcolor{customorange!80}\textbf{Probabilistic}}
& \multicolumn{2}{c}{\cellcolor{customorange!80}\textbf{Pointwise}}
& \multirow{2}{*}{\textbf{Rank}} \\
\cmidrule(lr){2-3}
\cmidrule(lr){4-5}
& \textbf{AQL}
& \textbf{AQCR}
& \textbf{MAE}
& \textbf{RMSE}
& \\
\midrule

No Mask
& 6.16 & 0.00 & 14.68 & 37.05 & 3 \\

Random
& 5.90 & 0.00 & 13.89 & 34.92 & 2 \\

Reverse
& 7.54 & 0.00 & 16.94 & 40.29 & 4 \\

\rowcolor{customgray!20}
Dual$^{\dagger}$
& \textbf{4.72} & \textbf{0.00} & \textbf{11.29} & \textbf{28.40} & \textbf{1} \\

\midrule

$k=0$
& 5.61 & 0.00 & 13.46 & 33.80 & 4 \\

$k=1$
& 4.84 & 0.00 & 12.02 & 29.22 & 3 \\

\rowcolor{customgray!20}
$k=2^{\dagger}$
& \textbf{4.72} & \textbf{0.00} & \textbf{11.29} & \textbf{28.40} & \textbf{1} \\

$k=4$
& 4.73 & 0.00 & 11.31 & 28.43 & 2 \\

\midrule

No Resid.
& 4.91 & 0.00 & 11.86 & 30.08 & 3 \\

Max Pool
& 4.96 & 0.00 & 11.98 & 30.36 & 4 \\

Concat.
& 4.74 & 0.00 & {11.27} & 28.42 & 2 \\

\rowcolor{customgray!20}
Residual$^{\dagger}$
& \textbf{4.72} & \textbf{0.00} & \textbf{11.29} & \textbf{28.40} & \textbf{1} \\

\midrule

Single-Q
& 4.73 & 2.07 & 11.31 & 28.39 & 4 \\

Multi-Q
& 4.72 & 1.97 & 11.30 & 28.44 & 3 \\

Post-Hoc
& 4.73 & 0.00 & 11.31 & {28.38} & 2 \\

\rowcolor{customgray!20}
Hier.$^{\dagger}$
& \textbf{4.72} & \textbf{0.00} & \textbf{11.29} & \textbf{28.40} & \textbf{1} \\

\bottomrule
\end{tabular}
\end{table}

\subsection{Aggregation Layer}

\begin{itemize}
    \item \textbf{No Residual Connection.}  
    Only the representations with the maximum degree of interactions from Eq.~\ref{eq:ResRail} are retained:
    \begin{equation}
        \mathbf{C}_i =  \mathbf{C}_{i,K}^{(+)} + \mathbf{C}_{i,K}^{(-)} .
        \label{eq:NoResRail}
    \end{equation}

    \item \textbf{Global Max Pooling.}  
    The global average pooling in Eq.~\ref{eq:avgpool} is replaced with global max pooling: 
    \begin{equation}
        \mathbf{U}_i = \max_{1 \leq j \leq T_{\max}} \mathbf{C}_i[j],
        \label{eq:maxpool}
    \end{equation}

    \item \textbf{Concatenation.}  The residual connection, as defined in Eq.~\ref{eq:ResRail}, is replaced with concatenation: 
    \begin{equation}
    \mathbf{C}_i = \sum_{k=1}^{K} \left( \mathbf{C}_{i,k}^{(+)} \| \mathbf{C}_{i,k}^{(-)} \right).
    \label{eq:res_concat}
\end{equation}
\end{itemize}

Results in Table~\ref{tab:ablation_study_compact} show that retaining only the representations with the maximum degree increases the AQL value by \textbf{4.03\%}, as this operation loses low-level features and leads to suboptimal performance.
Replacing the average pooling with max pooling leads to a performance drop of \textbf{5.08\%} in AQL. Given that the prediction targets are VWAPs, average pooling offers a useful inductive bias.
Moreover, replacing the residual connection with concatenation yields comparable results but introduces additional parameters, increasing the model complexity unnecessarily. Therefore, we recommend using the residual connection.

\subsection{Hierarchical Multi-Quantile Head}
\begin{itemize}
    \item \textbf{Single-Quantile Head.}  
    The proposed hierarchical head is replaced with a single-quantile head. Therefore, multiple models are trained independently for different quantiles.
    \item \textbf{Multi-Quantile Head.}  
    The hierarchical head is replaced with a standard multi-quantile head, where a single model produces multiple quantile forecasts using a shared representation.

    \item \textbf{Post-Hoc Sorting.}  
    The predicted quantiles made by individual models are reordered
    in ascending order.  
\end{itemize}

From Table~\ref{tab:ablation_study_compact}, we observe that the single-quantile model and multi-quantile model achieve comparable AQL and MAE values. However, these methods suffer from quantile crossing, with an AQCR of \textbf{2.07\%} and \textbf{1.97\%}, respectively. This issue is expected to become more pronounced as the number of predicted quantiles increases, since more quantile levels introduce more opportunities for monotonicity violations.
Although post-hoc sorting mitigates quantile crossing and achieves similar predictive performance, it introduces additional and unnecessary post-processing after model inference. This not only increases computational overhead but also complicates the modeling pipeline by separating quantile calibration from the main learning objective. In contrast, the designed hierarchical head enforces quantile monotonicity directly within the network architecture, eliminating quantile crossing while preserving a clean end-to-end design.


\section{Conclusion}
\label{sec:conclusion}
In this work, we develop a new methodology and open-source OrderFusion, 
an end-to-end and parameter-efficient model that consistently outperforms multiple baselines, generalizes across markets with both high (German) and low (Austrian) liquidity, \textcolor{black}{and overcomes the quantile crossing issue}. 
The results reveal that CID electricity markets do not exhibit perfect weak-form efficiency, highlighting the value of historical trades. Our findings further underscore the importance of explicitly modeling buy–sell interactions and injecting domain priors into models rather than solely stacking the model parameters.
Despite strong performance, several limitations remain. As with all forecasting models, the value of extra predictive variables remains open for consideration. Thus, as CID trading for sequential delivery periods occurs in parallel, incorporating trades from neighboring delivery periods, as well as periodic exogenous variables related to supply and demand conditions would be expected to further improve forecasting.
Nevertheless,  if CID markets evolve toward perfect weak-form efficiency, simple last-price models may suffice. 

\section*{Appendix}

\subsection{Hardware and Computation}
\label{appendix:hardware}
OrderFusion is evaluated on both an \textbf{NVIDIA A100 GPU} and an \textbf{Intel Core i7-1265U CPU}. The A100 targets high-performance computing, whereas the i7-1265U is a power-efficient processor used in standard laptops. 
Training requires approximately \textbf{1.5 minutes} on the A100 and \textbf{6 minutes} on the i7, while inference time is under \textbf{1 second} and \textbf{2.5 seconds}, respectively, making the model suitable for continuous use.

\subsection{Other Macro-Feature-Based Methods}
Several approaches have been used for intraday electricity price prediction, including Time2Vec-Transformer~\cite{cantillo2023intra}, Normalizing Flows~\cite{23Multivariate}, Long Short-Term Memory (LSTM)~\cite{Intraday}, Generative Model~\cite{chen2025probabilisticintradayelectricityprice}, and Bayesian Hierarchical Model~\cite{nickelsen2025bayesian}. However, these methods model price dynamics from exogenous fundamentals (e.g., load and renewable generation) and do not leverage the orderbook microstructure targeted in this work. 
Our experiments are therefore scoped to a microstructure-only setting: we use inputs that are derived from the  orderbook to quantify the predictive contribution of orderbook dynamics under a controlled design. This scope is complementary to fundamentals-based forecasting and should not be interpreted as claiming superiority in settings where rich exogenous information is available. Incorporating exogenous fundamentals alongside orderbook microstructure is straightforward within our framework and is an important direction for future work.

\subsection{Generalizability}
The proposed method targets the broader landscape of CID electricity markets operated by EPEX Spot. 
EPEX Spot’s services currently already span over {22} European countries, and an increasing number of countries are adopting CID markets to better manage short-term energy imbalances. 
Since the orderbook data format provided by EPEX Spot is standardized across regions, {OrderFusion} can be directly applied to multiple markets beyond the specific areas evaluated in this paper. 

\begin{table}[t]
\caption{Hyperparameter search space.}
\label{tab:hyperparams_models}
\begin{center}
\begin{tabular}{ll}
\toprule
\textbf{Model} & \textbf{Search Space} \\
\midrule

LQR & 
\begin{tabular}[t]{@{}l@{}}
\(\ell_1\): \{5e-8, 1e-8, 5e-7, 1e-7, \(\ldots\) , 1\}
\end{tabular}
\\
\midrule

MLP &
\begin{tabular}[t]{@{}l@{}}
\texttt{hidden\_size}: \{4, 16, 64, 256, 512\} \\
\texttt{n\_layers}: \{1, 2, 4\} \\
\texttt{dropout}: \{0.1, 0.2, 0.4\} \\
\end{tabular}
\\
\midrule

FEDFormer &
\begin{tabular}[t]{@{}l@{}}
\texttt{hidden\_size}: \{4, 16, 64, 256, 512\} \\
\texttt{conv\_hidden\_size}: \{8, 32, 128\} \\
\texttt{n\_layers}: \{1, 2, 4\} \\
\texttt{n\_heads}: \{1, 2, 4\} \\
\texttt{moving\_window}: \{4, 16, 64\} \\
\end{tabular}
\\
\midrule

iTransformer &
\begin{tabular}[t]{@{}l@{}}
\texttt{hidden\_size}: \{4, 16, 64, 256, 512\} \\
\texttt{n\_layers}: \{1, 2, 4\} \\
\texttt{n\_heads}: \{1, 2, 4\} \\
\texttt{d\_ff}: \{512, 1024, 2048\} \\
\texttt{dropout}: \{0.1, 0.2, 0.4\} \\
\end{tabular}
\\
\midrule

PatchTST &
\begin{tabular}[t]{@{}l@{}}
\texttt{hidden\_size}: \{4, 16, 64, 256, 512\} \\
\texttt{n\_layers}: \{1, 2, 4\} \\
\texttt{n\_heads}: \{1, 2, 4\} \\
\texttt{patch\_len}: \{4, 8, 16\} \\
\texttt{dropout}: \{0.1, 0.2, 0.4\} \\
\end{tabular}
\\
\midrule

TimesNet &
\begin{tabular}[t]{@{}l@{}}
\texttt{hidden\_size}: \{4, 16, 64, 256, 512\} \\
\texttt{conv\_hidden\_size}: \{8, 32, 128\} \\
\texttt{n\_layers}: \{1, 2, 4\} \\
\texttt{top\_k}: \{1, 2, 4\} \\
\end{tabular}
\\
\midrule

TimeXer &
\begin{tabular}[t]{@{}l@{}}
\texttt{hidden\_size}: \{4, 16, 64, 256, 512\} \\
\texttt{n\_layers}: \{1, 2, 4\} \\
\texttt{n\_heads}: \{1, 2, 4\} \\
\texttt{d\_ff}: \{64, 256, 1024\} \\
\end{tabular}
\\
\midrule
OrderFusion &
\begin{tabular}[t]{@{}l@{}}
\texttt{hidden\_size}: \{4, 16, 64, 256, 512\}  \\
\texttt{cutoff\_length}: \{\(2^0\), \(2^1\), \(2^2\), \(\ldots\) , \(2^{10}\)\} \\
\texttt{interaction\_degree}: \{1, 2, 4\} \\
\end{tabular}
\\
\bottomrule
\end{tabular}
\end{center}
\end{table}

\printbibliography[heading=bibintoc,title=Reference]

@misc{yu2025orderbookfeaturelearningasymmetric,
      title={Orderbook Feature Learning and Asymmetric Generalization in Intraday Electricity Markets}, 
      author={Runyao Yu and Ruochen Wu and Yongsheng Han and Jochen L. Cremer},
      year={2026},
      eprint={2510.12685},
      archivePrefix={arXiv},
      primaryClass={q-fin.CP},
      url={https://arxiv.org/abs/2510.12685}, 
}

@inproceedings{price_jumps,
  title={Price jumps on the intraday energy market - design and implementation of an alarm system with machine learning methods},
  author={Richard Lackes and Julian Sengewald and Mathis Wilz},
  booktitle={Pacific Asia Conference on Information Systems},
  year={2025},
  url={https://api.semanticscholar.org/CorpusID:280034961}
}

@article{iTrans,
  title={iTransformer: Inverted Transformers Are Effective for Time Series Forecasting},
  author={Liu, Yong and Hu, Tengge and Zhang, Haoran and Wu, Haixu and Wang, Shiyu and Ma, Lintao and Long, Mingsheng},
  journal={arXiv preprint arXiv:2310.06625},
  year={2023}
}

@inproceedings{PatchTST,
  title     = {A Time Series is Worth 64 Words: Long-term Forecasting with Transformers},
  author    = {Nie, Yuqi and
               H. Nguyen, Nam and
               Sinthong, Phanwadee and 
               Kalagnanam, Jayant},
  booktitle = {International Conference on Learning Representations},
  year      = {2023}
}

@inproceedings{xie2017aggregated,
  title={Aggregated residual transformations for deep neural networks},
  author={Xie, Saining and Girshick, Ross and Doll{\'a}r, Piotr and Tu, Zhuowen and He, Kaiming},
  booktitle={Proceedings of the IEEE conference on computer vision and pattern recognition},
  pages={1492--1500},
  year={2017}
}

@article{hasbrouck1991measuring,
  title={Measuring the information content of stock trades},
  author={Hasbrouck, Joel},
  journal={The Journal of Finance},
  volume={46},
  number={1},
  pages={179--207},
  year={1991},
  publisher={Wiley Online Library}
}

@book{hasbrouck2007empirical,
  title={Empirical market microstructure: The institutions, economics, and econometrics of securities trading},
  author={Hasbrouck, Joel},
  year={2007},
  publisher={Oxford University Press}
}

@article{bacry2015hawkes,
  title={Hawkes processes in finance},
  author={Bacry, Emmanuel and Mastromatteo, Iacopo and Muzy, Jean-Fran{\c{c}}ois},
  journal={Market Microstructure and Liquidity},
  volume={1},
  number={01},
  pages={1550005},
  year={2015},
  publisher={World Scientific}
}

@article{nagel1995unraveling,
  title={Unraveling in guessing games: An experimental study},
  author={Nagel, Rosemarie},
  journal={The American economic review},
  volume={85},
  number={5},
  pages={1313--1326},
  year={1995},
  publisher={JSTOR}
}

@article{fama1970efficient,
  title={Efficient capital markets: A review of theory and empirical work},
  author={Fama, Eugene F},
  journal={The journal of Finance},
  volume={25},
  number={2},
  pages={383--417},
  year={1970},
  publisher={JSTOR}
}

@article{LAGO2021116983,
title = {Forecasting day-ahead electricity prices: A review of state-of-the-art algorithms, best practices and an open-access benchmark},
journal = {Applied Energy},
volume = {293},
pages = {116983},
year = {2021},
issn = {0306-2619},
doi = {https://doi.org/10.1016/j.apenergy.2021.116983},
url = {https://www.sciencedirect.com/science/article/pii/S0306261921004529},
author = {Jesus Lago and Grzegorz Marcjasz and Bart {De Schutter} and Rafał Weron}
}

@article{koch2019short,
  title={Short-term electricity trading for system balancing: An empirical analysis of the role of intraday trading in balancing Germany's electricity system},
  author={Koch, Christopher and Hirth, Lion},
  journal={Renewable and Sustainable Energy Reviews},
  volume={113},
  pages={109275},
  year={2019},
  publisher={Elsevier}
}

@article{Probab,
  title={Probabilistic price forecasting for day-ahead and intraday markets: Beyond the statistical model},
  author={Andrade, Jos{\'e} R and Filipe, Jorge and Reis, Marisa and Bessa, Ricardo J},
  journal={Sustainability},
  volume={9},
  number={11},
  pages={1990},
  year={2017},
  publisher={MDPI}
}

@article{Forecasting,
  title={Forecasting the price distribution of continuous intraday electricity trading},
  author={Janke, Tim and Steinke, Florian},
  journal={Energies},
  volume={12},
  number={22},
  pages={4262},
  year={2019},
  publisher={MDPI}
}

@article{Ensemble,
  title={Ensemble forecasting for intraday electricity prices: Simulating trajectories},
  author={Narajewski, Micha{\l} and Ziel, Florian},
  journal={Applied Energy},
  volume={279},
  pages={115801},
  year={2020},
  publisher={Elsevier}
}

@article{Econometric,
  title={Econometric modelling and forecasting of intraday electricity prices},
  author={Narajewski, Micha{\l} and Ziel, Florian},
  journal={Journal of Commodity Markets},
  volume={19},
  pages={100107},
  year={2020},
  publisher={Elsevier}
}

@article{23Multivariate,
  title={Multivariate probabilistic forecasting of intraday electricity prices using normalizing flows},
  author={Cramer, Eike and Witthaut, Dirk and Mitsos, Alexander and Dahmen, Manuel},
  journal={Applied Energy},
  volume={346},
  pages={121370},
  year={2023},
  publisher={Elsevier}
}

@article{Trading,
  title={Trading on short-term path forecasts of intraday electricity prices},
  author={Serafin, Tomasz and Marcjasz, Grzegorz and Weron, Rafa{\l}},
  journal={Energy Economics},
  volume={112},
  pages={106125},
  year={2022},
  publisher={Elsevier}
}

@article{24Multivariate,
  title={Multivariate simulation-based forecasting for intraday power markets: Modeling cross-product price effects},
  author={Hirsch, Simon and Ziel, Florian},
  journal={Applied Stochastic Models in Business and Industry},
  volume={40},
  number={6},
  pages={1571--1595},
  year={2024},
  publisher={Wiley Online Library}
}

@article{Simulation,
  title={Simulation-based forecasting for intraday power markets: Modelling fundamental drivers for location, shape and scale of the price distribution},
  author={Hirsch, Simon and Ziel, Florian},
  journal={The Energy Journal},
  volume={45},
  number={3},
  pages={107--144},
  year={2024},
  publisher={SAGE Publications Sage CA: Los Angeles, CA}
}

@article{Understanding,
title = {Understanding intraday electricity markets: Variable selection and very short-term price forecasting using LASSO},
journal = {International Journal of Forecasting},
volume = {35},
number = {4},
pages = {1533-1547},
year = {2019},
issn = {0169-2070},
doi = {https://doi.org/10.1016/j.ijforecast.2019.02.001},
author = {Bartosz Uniejewski and Grzegorz Marcjasz and Rafał Weron},
}

@article{Short,
  title={Short-term price forecasting models based on artificial neural networks for intraday sessions in the Iberian electricity market},
  author={Monteiro, Claudio and Ramirez-Rosado, Ignacio J and Fernandez-Jimenez, L Alfredo and Conde, Pedro},
  journal={Energies},
  volume={9},
  number={9},
  pages={721},
  year={2016},
  publisher={MDPI}
}

@Article{beating,
AUTHOR = {Marcjasz, Grzegorz and Uniejewski, Bartosz and Weron, Rafał},
TITLE = {Beating the Naïve—Combining LASSO with Naïve Intraday Electricity Price Forecasts},
JOURNAL = {Energies},
VOLUME = {13},
YEAR = {2020},
NUMBER = {7},
ARTICLE-NUMBER = {1667},
URL = {https://www.mdpi.com/1996-1073/13/7/1667},
ISSN = {1996-1073},
DOI = {10.3390/en13071667}
}

@article{Intraday,
  title={Intraday electricity price forecasting via LSTM and trading strategy for the power market: A case study of the West Denmark DK1 grid region},
  author={K{\i}l{\i}{\c{c}}, Deniz Kenan and Nielsen, Peter and Thibbotuwawa, Amila},
  journal={Energies},
  volume={17},
  number={12},
  pages={2909},
  year={2024},
  publisher={MDPI}
}

@article{sortquantilecrossing1,
title = {A hybrid model for GEFCom2014 probabilistic electricity price forecasting},
journal = {International Journal of Forecasting},
volume = {32},
number = {3},
pages = {1051-1056},
year = {2016},
issn = {0169-2070},
doi = {https://doi.org/10.1016/j.ijforecast.2015.11.008},
author = {Katarzyna Maciejowska and Jakub Nowotarski},
}

@Article{sortquantilecrossing2,
AUTHOR = {Serafin, Tomasz and Uniejewski, Bartosz and Weron, Rafał},
TITLE = {Averaging Predictive Distributions Across Calibration Windows for Day-Ahead Electricity Price Forecasting},
JOURNAL = {Energies},
VOLUME = {12},
YEAR = {2019},
NUMBER = {13},
ARTICLE-NUMBER = {2561},
ISSN = {1996-1073},
DOI = {10.3390/en12132561}
}

@inproceedings{FEDFormer,
  title={Fedformer: Frequency enhanced decomposed transformer for long-term series forecasting},
  author={Zhou, Tian and Ma, Ziqing and Wen, Qingsong and Wang, Xue and Sun, Liang and Jin, Rong},
  booktitle={International conference on machine learning},
  pages={27268--27286},
  year={2022},
  organization={PMLR}
}

@inproceedings{
TimesNet,
title={TimesNet: Temporal 2D-Variation Modeling for General Time Series Analysis},
author={Haixu Wu and Tengge Hu and Yong Liu and Hang Zhou and Jianmin Wang and Mingsheng Long},
booktitle={The Eleventh International Conference on Learning Representations },
year={2023},
url={https://openreview.net/forum?id=ju_Uqw384Oq}
}

@inproceedings{TimeXer,
 author = {Wang, Yuxuan and Wu, Haixu and Dong, Jiaxiang and Qin, Guo and Zhang, Haoran and Liu, Yong and Qiu, Yunzhong and Wang, Jianmin and Long, Mingsheng},
 booktitle = {Advances in Neural Information Processing Systems},
 editor = {A. Globerson and L. Mackey and D. Belgrave and A. Fan and U. Paquet and J. Tomczak and C. Zhang},
 pages = {469--498},
 publisher = {Curran Associates, Inc.},
 title = {TimeXer: Empowering Transformers for Time Series Forecasting with Exogenous Variables},
 volume = {37},
 year = {2024}
}

@Article{drek,
AUTHOR = {Abramova, Ekaterina and Bunn, Derek},
TITLE = {Forecasting the Intra-Day Spread Densities of Electricity Prices},
JOURNAL = {Energies},
VOLUME = {13},
YEAR = {2020},
NUMBER = {3},
ARTICLE-NUMBER = {687},
URL = {https://www.mdpi.com/1996-1073/13/3/687},
ISSN = {1996-1073},
DOI = {10.3390/en13030687}
}

@ARTICLE{variance_stab,
  author={Uniejewski, Bartosz and Weron, Rafał and Ziel, Florian},
  journal={IEEE Transactions on Power Systems}, 
  title={Variance Stabilizing Transformations for Electricity Spot Price Forecasting}, 
  year={2018},
  volume={33},
  number={2},
  pages={2219-2229},
  doi={10.1109/TPWRS.2017.2734563}}

@article{nickelsen2025bayesian,
  title={Bayesian hierarchical probabilistic forecasting of intraday electricity prices},
  author={Nickelsen, Daniel and M{\"u}ller, Gernot},
  journal={Applied Energy},
  volume={380},
  pages={124975},
  year={2025},
  publisher={Elsevier}
}

@ARTICLE{9788043,
  author={Sgarlato, Raffaele and Ziel, Florian},
  journal={IEEE Transactions on Power Systems}, 
  title={The Role of Weather Predictions in Electricity Price Forecasting Beyond the Day-Ahead Horizon}, 
  year={2023},
  volume={38},
  number={3},
  pages={2500-2511},
  doi={10.1109/TPWRS.2022.3180119}}

@misc{yu2026pricefmfoundationmodelprobabilistic,
      title={PriceFM: Foundation Model for Probabilistic Electricity Price Forecasting}, 
      author={Runyao Yu and Chenhui Gu and Jochen Stiasny and Qingsong Wen and Wasim Sarwar Dilov and Lianlian Qi and Jochen L. Cremer},
      year={2026},
      eprint={2508.04875},
      archivePrefix={arXiv},
      primaryClass={cs.CE},
      url={https://arxiv.org/abs/2508.04875}, 
}

@ARTICLE{10012043,
  author={Zhang, Chenxu and Fu, Yong},
  journal={IEEE Transactions on Power Systems}, 
  title={Probabilistic Electricity Price Forecast With Optimal Prediction Interval}, 
  year={2024},
  volume={39},
  number={1},
  pages={442-452},
  doi={10.1109/TPWRS.2023.3235193}}

@ARTICLE{11045523,
  author={Su, Heng-Yi and Liao, Guan-Zhang},
  journal={IEEE Transactions on Power Systems}, 
  title={A Spike-Resilient Temporal-Adaptive Neural Framework for Day-Ahead Electricity Price Interval Forecasting}, 
  year={2025},
  volume={40},
  number={5},
  pages={4411-4414},
  doi={10.1109/TPWRS.2025.3581805}}

@misc{yu2026deeplearningelectricityprice,
      title={Deep Learning for Electricity Price Forecasting: A Review of Day-Ahead, Intraday, and Balancing Electricity Markets}, 
      author={Runyao Yu and Derek W. Bunn and Julia Lin and Jochen Stiasny and Fabian Leimgruber and Tara Esterl and Yuchen Tao and Lianlian Qi and Yujie Chen and Wentao Wang and Jochen L. Cremer},
      year={2026},
      eprint={2602.10071},
      archivePrefix={arXiv},
      primaryClass={q-fin.CP},
      url={https://arxiv.org/abs/2602.10071}, 
}

@article{SCHARFF2016544,
title = {Trading behaviour on the continuous intraday market Elbas},
journal = {Energy Policy},
volume = {88},
pages = {544-557},
year = {2016},
issn = {0301-4215},
doi = {https://doi.org/10.1016/j.enpol.2015.10.045},
url = {https://www.sciencedirect.com/science/article/pii/S0301421515301713},
author = {Richard Scharff and Mikael Amelin}
}

@article{LEHNA2022105742,
title = {Forecasting day-ahead electricity prices: A comparison of time series and neural network models taking external regressors into account},
journal = {Energy Economics},
volume = {106},
pages = {105742},
year = {2022},
issn = {0140-9883},
doi = {https://doi.org/10.1016/j.eneco.2021.105742},
url = {https://www.sciencedirect.com/science/article/pii/S0140988321005879},
author = {Malte Lehna and Fabian Scheller and Helmut Herwartz}
}

@inproceedings{brando2022deep,
  title={Deep non-crossing quantiles through the partial derivative},
  author={Brando, Axel and Center, Barcelona Supercomputing and Rodr{\'\i}guez-Serrano, Jos{\'e} and Vitri{\`a}, Jordi and others},
  booktitle={International Conference on Artificial Intelligence and Statistics},
  pages={7902--7914},
  year={2022},
  organization={PMLR}
}

@inproceedings{monteiro2022monotonicity,
  title={Monotonicity regularization: Improved penalties and novel applications to disentangled representation learning and robust classification},
  author={Monteiro, Joao and Ahmed, Mohamed Osama and Hajimirsadeghi, Hoseein and Mori, Greg},
  booktitle={Uncertainty in Artificial Intelligence},
  pages={1381--1391},
  year={2022},
  organization={PMLR}
}

@article{cantillo2023intra,
  title={An intra-day electricity price forecasting based on a probabilistic transformer neural network architecture},
  author={Cantillo-Luna, Sergio and Moreno-Chuquen, Ricardo and Lopez-Sotelo, Jesus and Celeita, David},
  journal={Energies},
  volume={16},
  number={19},
  pages={6767},
  year={2023},
  publisher={MDPI}
}

@misc{chen2025probabilisticintradayelectricityprice,
      title={Probabilistic intraday electricity price forecasting using generative machine learning}, 
      author={Jieyu Chen and Sebastian Lerch and Melanie Schienle and Tomasz Serafin and Rafał Weron},
      year={2025},
      eprint={2506.00044},
      archivePrefix={arXiv},
      primaryClass={stat.AP},
      url={https://arxiv.org/abs/2506.00044}, 
}
\end{document}